\documentclass[%
aip,
amsmath,amssymb,
reprint,%
]{revtex4-2}

\usepackage{framed,multirow}
\usepackage[pdftex]{graphicx}
\usepackage[T2A]{fontenc}
\usepackage[utf8]{inputenc}

\usepackage[russian,english]{babel}
\usepackage{amssymb}
\usepackage{latexsym}

\usepackage{amsmath}
\usepackage{cmap}
\usepackage{multirow}
\usepackage{layouts}

\usepackage{caption}
\usepackage{float}
\usepackage{placeins}
\usepackage{hyperref}

\hypersetup{				
	colorlinks=true,       	
	linkcolor=blue,          
	citecolor=blue,        
}

\usepackage[usenames]{color}

\begin{document}

\title{Drift-kinetic PIC model for simulations of longitudinal plasma confinement in mirror traps}

\author{V.V. Glinskiy}
\affiliation{Budker Institute of Nuclear Physics SB RAS, 630090, Novosibirsk, Russia}
\affiliation{Novosibirsk State University, 630090, Novosibirsk, Russia}
\email{v.v.glinskiy@yandex.ru}
\author{I.V. Timofeev}
\affiliation{Budker Institute of Nuclear Physics SB RAS, 630090, Novosibirsk, Russia}
\affiliation{Novosibirsk State University, 630090, Novosibirsk, Russia}
\author{V.V. Prikhodko}
\affiliation{Budker Institute of Nuclear Physics SB RAS, 630090, Novosibirsk, Russia}

\begin{abstract}

The paper presents a 1D2V electrostatic PIC model with a drift-kinetic description of all particle types aiming at simulating classical longitudinal plasma transport in axially symmetric open traps. The model generalizes the semi-implicit particle-in-cell method with exact conservation of energy and charge to the case of collisional plasma and adapts it to boundary conditions on perfectly conducting walls with a floating potential. Implementation of Coulomb collisions is tested on the problem of temperature relaxation in a two-component plasma and demonstrates good agreement with the analytical theory. Since quasi-neutrality of plasma is not strictly determined, the model is able to correctly reproduce the ambipolar electric potential profile up to the walls. At the same time, the main advantage of implicit PIC simulations --- the ability to use large grid steps, many times larger than the Debye radius --- does not prevent the correct modeling of the near-wall electric potential jump. The model satisfactorily reproduces the known results of the Debye sheath theory and the Bohm criterion. A comparison of stationary plasma profiles formed in a mirror trap in the presence of a constant particle source with the results of simulations using the hybrid code MIDAS showed that self-consistent consideration of electron kinetics in expanders leads to noticeable (at the level of 15\%) differences in the electron temperature, potential, and density of the confined plasma.   

\end{abstract}

\maketitle

\section{Introduction}

Modeling the processes of plasma confinement in magnetic traps is one of the most pressing and computationally challenging problems in plasma physics. Growing interest in magnetic traps with open field lines has recently stimulated the development of numerical models capable of taking into account the main feature of these systems --- direct contact of the plasma with the material walls. To weaken such contact, open traps use expanders in which the magnetic field is reduced by tens and hundreds of times. The challenge in modeling such systems lies in the need to kinetically describe not only the ions but also the electrons, which is extremely expensive due to the huge gap in the characteristic spatial and temporal scales determined by these types of particles. To overcome this difficulty, a hybrid approach is often used in simulations. While ions are described kinetically, electrons are replaced by a quasi-neutral fluid with a Boltzmann distribution and a uniform temperature along the whole system \cite{prikhodko2025numerical,caneses2025particle,Dorf2025}. This simplification may be justified in the central region between magnetic mirrors, where the electron component is close to thermodynamic equilibrium, but it becomes inapplicable in the expander where electrons are weakly collisional, meaning that their distribution function can deviate significantly from the Maxwellian \cite{wetherton2021drift,francisquez2023toward,Tyushev2025}. This does not allow the hybrid approach to correctly describe the physics of the expander and reliably predict the profile of the ambipolar potential which arises due to the escape of fast electrons from the trap and leads to the locking of the energy losses of the hot plasma on the wall \cite{ryutov2005axial,ivanov2017gas,soldatkina2020measurements}.

In the recent paper \cite{glinskiy20241d}, we have proposed a collisionless version of the one-dimensional drift-kinetic model  based on the energy-conserving semi-implicit particle-in-cell (PIC) method \cite{lapenta2017exactly, berendeev2024energy}. The main advantage of this semi-implicit model is its ability to kinetically describe all plasma components at large spatial steps (many times greater than the Debye radius $\lambda_D$) without the nonlinear iterations inherent in fully implicit schemes \cite{jimenez2024implicit,chen2023an}. Conservation of energy in such a numerical scheme is achieved because the electric field is found not from the Poisson equation (as usual for electrostatic models), but from the Ampere's law. In this case, for automatic implementation of the Gauss law (local conservation of charge), the current is corrected in the model in order to accurately satisfy the continuity equation.
Accurate local charge conservation provides lower noise levels, allowing for savings in the number of particles per cell. Energy conservation is also important for modeling long-term processes such as plasma confinement in real experimental setups.

However, to carry out a realistic numerical experiment aimed at studying classical longitudinal losses in experimental facilities such as GDT, GOL-NB or GDMT \cite{bagryansky2024progress,skovorodin2023gas,Sidorov2024}, it is necessary to add at least two key elements to the developed model: (i) Coulomb collisions between particles and (ii) a correct description of the interaction of plasma with the conducting walls limiting the plasma volume. Coulomb collisions are responsible for the scattering of ions into the loss cone and determine the plasma confinement time, while realistic boundary conditions allow the ambipolar potential and edge plasma to be correctly reproduced in the model. Thus, taking these two elements into account simultaneously is a necessary step to create a self-consistent model capable of quantitatively predicting the parameters of the confined plasma.

This paper presents the results of testing two modules that have been added to the drift-kinetic PIC code ADEPT (Axial Drift-kinetic Electrostatic code for Plasma Transport). The first module implements a Monte Carlo algorithm for energy-conserving binary Coulomb collisions  based on the Takizuka-Abe method \cite{takizuka1977binary}. The second module describes the absorption of particles by perfectly conducting walls, inside which the electric field must become zero. Similar boundary conditions have already been used in the previous paper \cite{glinskiy20241d}, but a detailed recipe for their implementation without violating the law of conservation of energy will be discussed for the first time.

To demonstrate the capabilities of the extended code, we simulate the problem of continuous injection of particles into a mirror trap with conductive end walls. Stationary profiles of density, temperature and other plasma parameters are compared with similar calculations using the hybrid code MIDAS \cite{prikhodko2025numerical}. This made it possible to determine the differences that the kinetic description of electrons leads to in this problem.

\section{1D drift-kinetic PIC code}

The numerical code ADEPT presented in this paper is based on the 1D2V PIC model \cite{glinskiy20241d}, in which the semi-implicit particle-in-cell method with exact conservation of energy and local charge is generalized to drift-kinetic equations that track the motions of not the particles themselves, but their Larmor centers. Unlike most electrostatic models, the electrostatic version of Ampère's law is used here instead of Poisson's equation to determine the electric field. In this paper, we show how to include Coulomb collisions in this model and implement perfectly conducting walls at the boundary of the region without violating the law of energy conservation. A key improvement to the code, which made it possible to conduct numerical experiments on a realistic scale, was the porting of the parallel CPU version of the code to graphics accelerators (GPUs). This resulted in significant performance gains. For example, testing on an NVIDIA Tesla V100 GPU demonstrated a 3-5x acceleration compared to performance on an AMD EPYC 7773X. The achieved computational speed made it possible to carry out simulations with the real mass of ions in a reasonable time.

Next, we describe in detail the main modules of the improved code. Section \ref{sec_kernel} is devoted to the basic algorithm described in \cite{glinskiy20241d}. Section \ref{sec_collisions} presents the implementation of Coulomb collisions. Finally, section \ref{sec_boundary_condition} discusses the algorithm for particle interactions with conducting walls.

\subsection{The kernel of the code} \label{sec_kernel}

In the drift-kinetic approximation, the longitudinal motion of the guiding centers of particles inside a thin magnetic field tube is described by the equations:
\begin{align}
	&\frac{d z_p}{dt}=v_{\| p}, \\
	&m_p\frac{dv_{\| p}}{dt}=q_p E(t,z_p(t))-\mu_p \left(\frac{d B}{dz}\right)_{z=z_p}, \\
	&\mu_p=\frac{m_p v_{\bot p}^2(t)}{2 B(z_p(t))}=\rm{const},
\end{align}
where $\mu_p$ is the magnetic moment of a particle that is conserved during its motion, $B(z)$ is the stationary magnetic field inside the tube, $v_\parallel$ and $v_\perp$ are  longitudinal and transverse to the magnetic field components of the particle velocity. Further in this section we will omit the index $\parallel$ and use dimensionless quantities that are measured in the following units: masses in the mass of an electron $m_e$, charges in the charge of an electron $e$, velocities in the speed of light $c$, time in the reciprocal plasma frequency $\omega_{pe}=\sqrt{4\pi e^2 n_0/m_e}$, spatial coordinates in $c/\omega_{pe}$, electric and magnetic fields in $m_e c \omega_{pe}/e$, current densities in $en_0 c$.

All particles of the same type, regardless of their position on the field line, have the same mass and charge $m_p=q_p=n_0 V_0/N_0$, where $n_0$ is the plasma density at the center of the system, $V_0$ is the volume of the central cell, and $N_0$ is the number of macroparticles in this cell. Thus, to set a uniform density in cells with a different grid coordinate $z_g=gh$ and a different volume $V_g$, we change the number of macroparticles $N_g=N_0 V_g/V_0$.
The shape of macroparticles is described in our model by the linear PIC kernel $W(z-z_p)$. Using this shape function, the grid values of the electric field $E_g$ are interpolated onto a particle with coordinate $z_p$ according to the rule
\begin{equation}
	E_p=\sum\limits_g E_g W(z_g-z_p),
\end{equation}
and the current from the particle with number $p$ is distributed among the grid nodes according to the formula
\begin{equation}
	J_g^p=q_p v_p W(z_g-z_p)/V_g.
\end{equation}

In our method, each time step consists of a prediction stage and a correction stage.
The prediction stage begins with a shift of a particle with the known velocity, $v_p^n$, by half a time step, $\tau/2$:
\begin{equation} \label{eq_first_push}
	z_p^{n+1/2}=z_p^{n}+\tau v_p^{n}/2,
\end{equation}
then the field values at the new time step $E_g^{n+1}$ are predicted by solving a system of linear algebraic equations \cite{glinskiy20241d}:
\begin{align}
	&E_g^{n+1}+\sum\limits_{g^\prime} \mathcal{L}_{g g^\prime} E_{g^\prime}^{n+1}=F_g, \label{eq_find_E}\\
	&\mathcal{L}_{g g^\prime}=\frac{\tau^2}{4 V_g}\sum\limits_p \frac{q_p^2}{m_p} W(z_g-z_p^{n+1/2}) W(z_{g^\prime}-z_p^{n+1/2}), \\
	&F_g=E_g^n- \frac{\tau}{V_g}\sum\limits_p q_p\times\nonumber\\ 
	&\times  \left[v_p^n+\frac{\tau}{2 m_p}\left(\frac{q_p}{2}E_p^n-\mu_p B_p^{\prime}\right)\right]W(z_g-z_p^{n+1/2}).
\end{align}
The predicted field is used to find the new coordinate of the particle:
\begin{equation}
	z_p^{n+1}=z_p^{n+1/2}+\tau v_p^{n+1}/2,
\end{equation}
where the velocities $v_p^{n+1}$ at a new time step are determined from the equation of motion
\begin{align}
	&m_p\frac{v_p^{n+1}-v_p^n}{\tau}=q_p E_p^{n+1/2}-\mu_p B_p^{\prime}, \label{eq_motion}\\
	&E_p^{n+1/2}=\sum\limits_{g}\frac{\left(E_g^{n+1}+E_g^{n}\right)}{2} W(z_g-z_p^{n+1/2}), \\
	&B_p^{\prime}=\sum\limits_{g} \frac{B_{g+1}-B_{g-1}}{2h} W(z_g-z_p^{n+1/2}).
\end{align}

\begin{figure*}
	\centering{\includegraphics[width=0.8\linewidth]{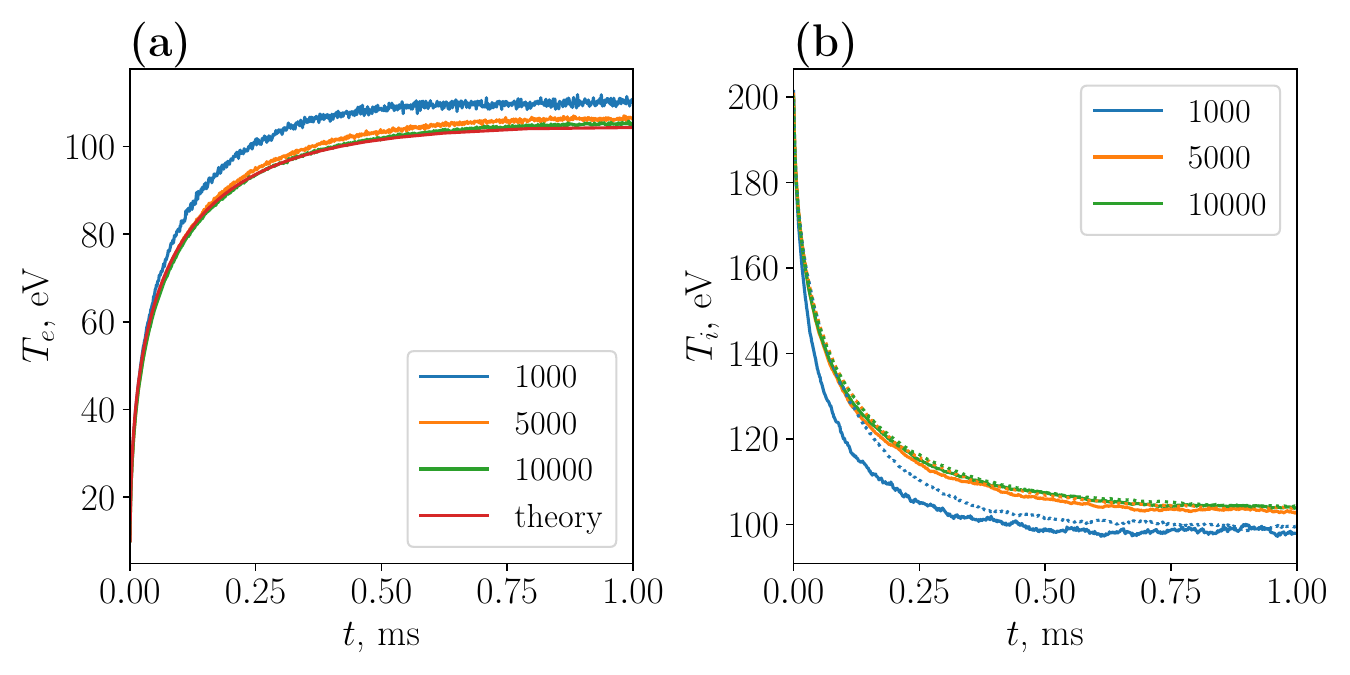}}
	\caption{Time dependence of the (a) electron and (b) ion temperatures averaged over the entire system for different numbers of particles in the cell. In (b), the solid curves show the longitudinal ion temperature $T_{i\parallel}$, while the dotted curves show the transverse $T_{i\perp}$. $T_{i\parallel}$ is the average doubled longitudinal kinetic energy $<mv^2_\parallel>$ of ions, and $T_{i\perp}$ is their average transverse energy $<mv^2_\perp/2>$. The red curve is the theoretical dependence of $T_e$ on time \eqref{eq_time}.}
	\label{fig_collision}
\end{figure*}

Since the current $\tilde{J}_g^{n+1/2}$ calculated from the particle shape does not satisfy the continuity equation exactly (or, equivalently, slightly violates the Gauss's law), and the chosen method of approximating the magnetic field gradient leads to a small error in the energy conservation law $Q$,
\begin{widetext}
\begin{equation}
	Q=\sum\limits_{p\in s} \mu_p \sum\limits_g \left\{\tau \tilde{v}_p^{n+1/2} \left(\frac{B_{g+1}-B_{g-1}}{2h}\right) W(z_g-z_p^{n+1/2})- B_g \left[W(z_g-z_p^{n+1})-W(z_g-z_p^{n})\right]\right\},
\end{equation}
\end{widetext}
a correction step is required in which the conservation of energy and charge will be restored. For this purpose, using the known positions of the particles $z_p^n$ and $z_p^{n+1}$, the current $J_g^{n+1/2}$ is calculated directly from the continuity equation using the density decomposition method \cite{esirkepov2001exact}. The electric field is then corrected
\begin{equation}
	E_g^{n+1}-\tilde{E}_g^{n+1}=-\tau \left(J_g^{n+1/2}-\tilde{J}_g^{n+1/2}\right),
	\end{equation}
which creates an additional error in energy.	To restore the law of global energy conservation, the particle velocities $\tilde{v}_{p}^{n+1}$ obtained at the prediction stage are multiplied by a correction coefficient, $v_{p}^{n+1}=\lambda \tilde{v}_{p}^{n+1}$,
	\begin{multline}
		\lambda^2 = 1+ \\
		\frac{Q+ \frac{\tau}{2}\sum\limits_g V_g \left(E_g^{n+1}+\tilde{E}_g^{n+1}\right) \left[J_{g}^{n+1/2}-\tilde{J}_{g}^{n+1/2}\right]}{\sum\limits_p m_p \left(\tilde{v}_p^{n+1}\right)^2/2}.
	\end{multline}

\subsection{Coulomb collisions} \label{sec_collisions}

To account for Coulomb collisions, a Monte Carlo algorithm was added to the code,
in which the well-known Takizuka-Abe method \cite{takizuka1977binary} is adapted to the two-dimensional velocity space $(v_{\bot},v_{\|})$ used in the drift-kinetic model.
The collision algorithm is executed either at every time step or several steps before the prediction stage of the main computational algorithm \ref{sec_kernel}, ensuring correct energy conservation in the system. The collision procedure is organized as follows.
The particles are sorted into cells, then paired within each cell. A sequence of transformations is then performed on each pair:
	\begin{itemize}
		\item transition to three-dimensional Cartesian velocities
		\begin{align*}
			v_x &= -\text{sgn}(q_p)v_{\perp}\cos\varphi, \\
			v_y &= \text{sgn}(q_p)v_{\perp}\sin\varphi, \\
			v_z &= v_{\parallel},
		\end{align*}
		where $sgn(q_p)$ is the sign of the charge of the colliding particle, $\varphi$ is a random angle;
		\item implementation of collisions using the Takizuka-Abe method \cite{takizuka1977binary};
		\item inverse transformation of Cartesian velocities into polar coordinates:
		\begin{align*}
			v_{\parallel} = v_z, \quad v_{\perp} = \sqrt{v_x^2 + v_y^2}.
		\end{align*}
	\end{itemize}

To verify the correct operation of Coulomb collisions, the problem of temperature equalization between electrons and ions in a two-component plasma was simulated. According to the analytical solution of this problem \cite{Kogan1958}, the electron heating time from the initial temperature $T_e(0)$ to the current value $T_e(t)$ is determined by the expression:
\begin{equation} \label{eq_time}
	t = \tau_{col} \int \limits^{T_e(t)/T_\infty}_{T_e(0)/T_\infty} \frac{[2/M + y(1-1/M)]^{3/2}}{1-y}dy,
\end{equation} 
where
\begin{align}
	&\tau_{col} = \frac{3}{4} \frac{(2\pi)^{3/2}}{\Lambda} \frac{n_ec^3}{\omega_{pe}^3} M \left( \frac{T_\infty}{m_ec^2} \right)^{3/2}, \\ 
	&T_\infty = \frac{T_e(0) + T_i(0)}{2}, \quad M = \frac{m_i}{m_e},  
\end{align}
$\omega_{pe} = \sqrt{4\pi n_e e^2/m_e}$, $m_e$, $n_e$ are the mass and density of electrons, and $m_i$ is the ion mass (from here on, ions mean protons).

In simulations carried out with periodic boundary conditions, at the initial moment of time, electrons and ions are uniformly distributed over a spatial grid with a size of 57 cells and have a Maxwellian velocity distribution with temperatures of $T_e(0) = 10$ eV and $T_i(0) = 200$ eV. The following parameters are fixed: Coulomb logarithm $\Lambda = 15$, plasma density $n = 5\cdot 10^{13}$ cm$^{-3}$, temporal and spatial grid step $h = 0.17$ cm, $\tau = 2.25 \cdot 10^{-11}$ s (with collisions occurring once every 25 time steps). The magnetic field is set uniform throughout the computational domain.

Analysis of the simulation results revealed a noticeable influence of numerical noise on relaxation processes. As can be seen from Fig. \ref{fig_collision}, in the case of using 1000 particles per cell, an acceleration of energy exchange between plasma components is observed compared to the theoretical prediction \eqref{eq_time}. This effect is associated with the numerical collisions caused by the presence of electric field noise. Since in the collisionless case such parasitic energy exchange occurs only between the longitudinal degrees of freedom of the particles, the transverse velocities are affected by this noise only indirectly through Coulomb collisions. For this reason, the main effect of the numerical collisions in our model is the ion temperature anisotropy observed in Fig. \ref{fig_collision} (b). Increasing the number of macroparticles in the cell to 5000 and higher makes it possible to significantly suppress this effect, which confirms its numerical nature. In this case, the distribution of ion temperature becomes isotropic, and the rate of energy exchange between electrons and ions agrees well with the prediction of analytical theory. This does not mean, however, that in all simulations the number of macroparticles in a cell should not fall below 5000. In systems with open boundary conditions, which are used to model mirror traps, the lifetime of particles can be much shorter than the time of energy exchange between components, so the integral effect of noise fields on these particles will remain small even at higher noise levels.

\subsection{Boundary conditions} \label{sec_boundary_condition}

\begin{figure}[htb]
	\centering
	\includegraphics[width=0.6\linewidth]{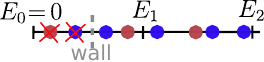}
	\caption{Implementation of open boundary conditions.}
	\label{fig_grid_wall}
\end{figure}

To correctly describe the formation of an ambipolar potential in a magnetic trap, we implement boundary conditions that simulate perfectly conducting walls located at the ends of the computational domain. The wall's position relative to the spatial grid nodes and the particle removal method are schematically depicted in Fig. \ref{fig_grid_wall}.
The numerical implementation of such boundaries requires the following conditions to be met. First, each particle that reaches any wall located at a distance of $h/2$ from the boundaries of the computational domain at each half-step in time is removed. Secondly, no current correction is performed at the two outermost nodes on each side of the system, which avoids taking into account the current contributions from the removed particles. Thirdly, the electric field at grid nodes inside the wall  is set to zero. Fourthly, the magnetic field at the three outermost nodes on each side of the system is assumed to be uniform.

\begin{figure}[htb]
	\centering
	\includegraphics[width=0.9\linewidth]{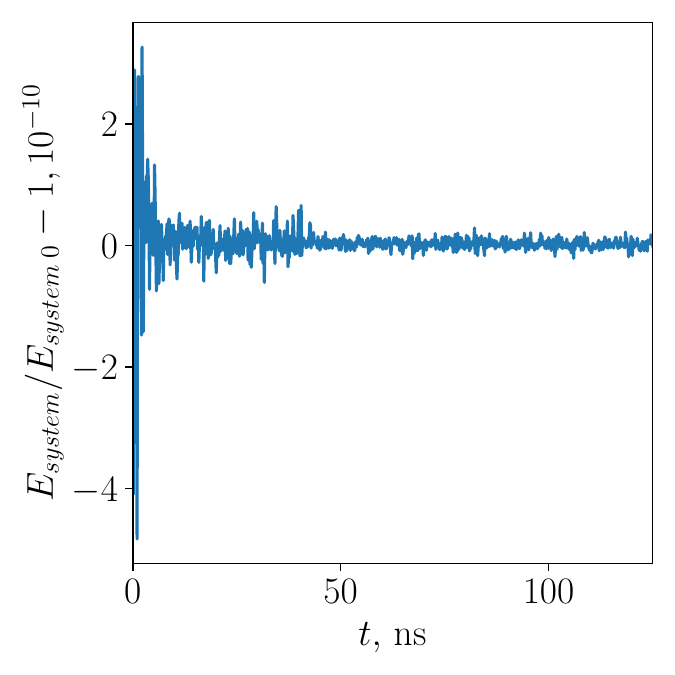}
	\caption{The energy of the system $E_{system} = E_p + E_{field} + E_{deleted\; p} - E_{added\;p}$ in the simulation with h=0.01 cm shown in Fig. 4  in units of its initial value $E_{system\;0}$ ($E_p$ is the energy of particles located inside the calculation region, $E_{field}$ is the energy of the electric field, $E_{deleted\; p}$ and $E_{added\;p}$ are the energies of the removed and injected particles).}
	\label{fig3}
\end{figure}

Since the direct zeroing of the electric field at the boundaries of the system after its determination from the equation \eqref{eq_find_E} leads to a violation of energy conservation in finite-difference form, we achieve the conversion of this field to zero by zeroing all local currents from particles in Ampere's law, that is, by the following change of the matrix $\mathcal{L}_{g g^\prime}$ and the vector $F_g$:
\begin{align}
	&\mathcal{L}_{0\;0} = \mathcal{L}_{0\;1} = \mathcal{L}_{1\;0} = \mathcal{L}_{N-1\;N-1} = \mathcal{L}_{N-1\;N-2} =\nonumber \\
	&= \mathcal{L}_{N-2\;N-1} = F_0 = F_{N-1} = 0,
\end{align}
where $N$ is a full number of grid nodes. As simulations show, this method of specifying boundary conditions leaves the matrix $\mathcal{L}$ symmetrical and does not violate the law of energy conservation in the system throughout the entire simulation time (Fig. \ref{fig3}).

We tested the correctness of the implemented boundary conditions by considering the formation of a charged layer near a wall with a floating potential. Similar to the work \cite{gyergyek2014potential}, for comparison with the Debye sheath theory, a stationary plasma with cold ions and hot electrons is maintained by continuously injecting new particles into a region bounded by two conducting walls. An anomalously high frequency of electron-electron collisions is used to maintain a Maxwellian electron velocity distribution.

\begin{figure*}
	\centering{\includegraphics[width=0.75\linewidth]{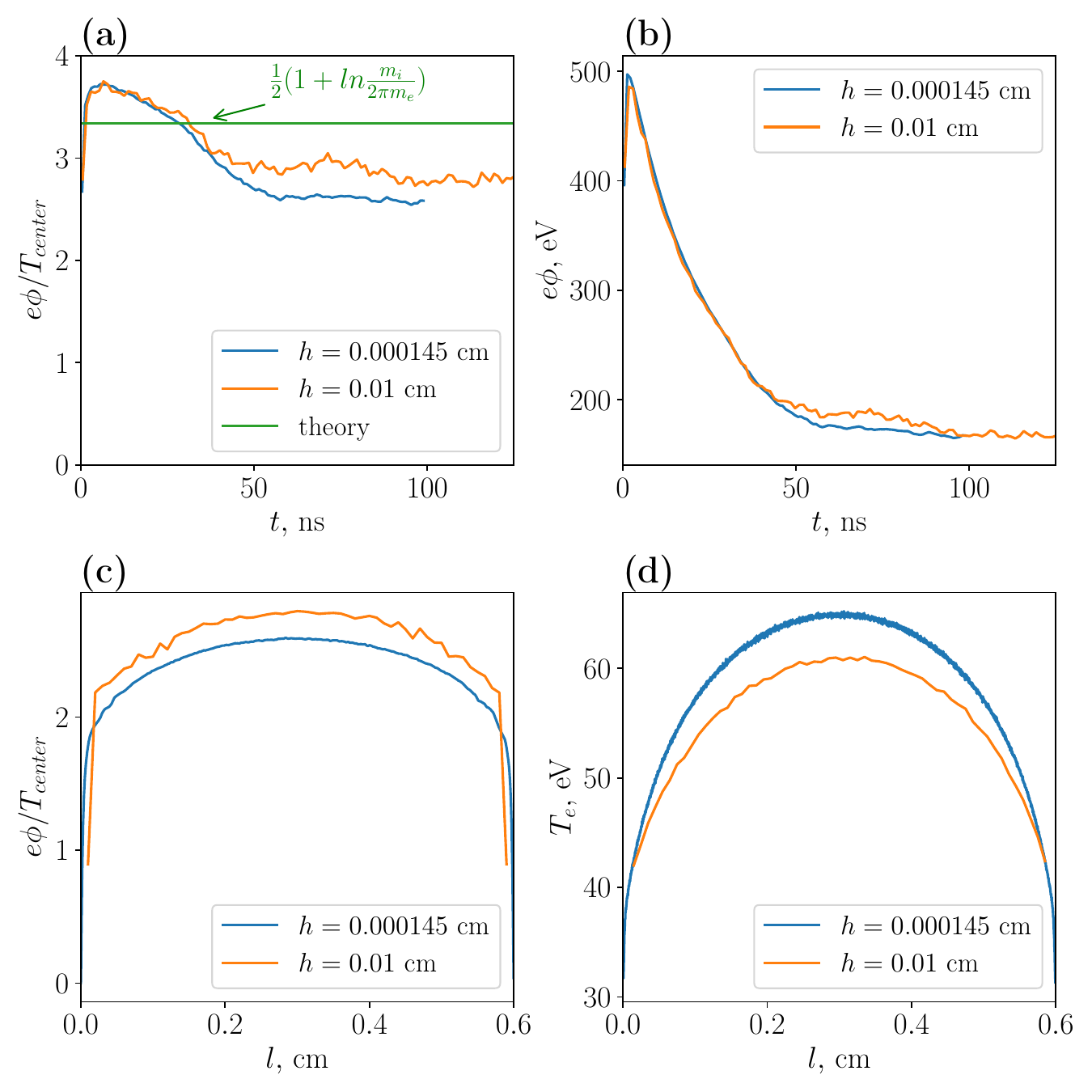}}
	\caption{Results of PIC simulations with different spatial steps $h = 0.000145$ cm (blue curve) and $h = 0.01$ cm (orange curve): (a) dependence of the electric potential jump between the system center and the wall on time in units of the central electron temperature $T_{center}/e$; (b) time behavior of the electric potential jump in electron volts, (c) spatial distribution of the potential $e\phi/T_{center}$ in the steady state ($t=100$ ns); (d) electron temperature profiles. All graphs are time averaged.}
	\label{fig_boundary_potential}
\end{figure*}

In our simulations, at the initial moment of time, the plasma with density $n_0 = 10^{13}$ cm$^{-3}$ is set to be uniform along the entire system, and the electrons and ions have temperatures $T_{e0} = 150$ eV and $T_{i0} = 0.1$ eV.
At subsequent times, electron-ion pairs are injected into the system at a rate of $dn_{\rm inject}/dt = 4.04\cdot 10^{20}$ cm$^{-3}/$s and the same temperatures. Anomalous electron scattering is achieved by incorporating Coulomb collisions with a greatly increased Coulomb logarithm $\Lambda = 15\cdot 10^5$. These collisions are performed exclusively for electron-electron pairs. Since the simulation of plasma confinement on the real scale of plasma experiments will have to be carried out with grid steps significantly exceeding the Debye radius $\lambda_D$, and the drop in electric potential in the charged Debye sheath by several electron temperatures \cite{ryutov2005axial,skovorodin2019suppression} is an important factor in electron confinement, it is necessary to find out how strongly the spatial resolution affects the formation of the potential jump in the near-wall layer. Two simulations of the same physical system are carried out for different $h$: in the first simulation, the spatial step $h = 0.000145$ cm ($\tau = 2.2\cdot 10^{-14}$ s, 3000 particles per cell) resolves well the Debye radius $\lambda_D \sim 0.0014$ cm, while in the second, the grid step $h = 0.01$ cm ($\tau = 1.5\cdot 10^{-12}$ s, 3000 particles per cell) does not allow for a detailed description of the spatial structure of the potential inside the near-wall layer. In theory, the electric potential of an undisturbed plasma relative to a floating wall is given by the expression:
\begin{equation} \label{eq_potential}
	e\phi_{plasma} = \frac{T_{e\; plasma}}{2}\left[1 + \ln\left(\frac{m_i}{2\pi m_e}\right)\right],
\end{equation}
where $T_{e\; plasma}$ is the electron temperature in an undisturbed plasma, and $e$ is the absolute value of the electron charge. The first term here describes the potential difference between the undisturbed plasma and the beginning of the charged layer (quasineutral pre-sheath), and the second one describes the potential jump inside the charged layer (see \cite{Lochte1968plasma}).

The electric potential profiles, as well as the time behavior of its jump between the wall and the plasma center in PIC simulations with different spatial resolutions, are shown in Figure \ref{fig_boundary_potential}. From Fig. \ref{fig_boundary_potential} (b) it is clear that the total potential jump is practically independent of whether the Debye scale is resolved or not. This agreement indicates that the losses of fast (above-barrier) electrons in both simulations are the same. If, for comparison with theory, we consider this potential jump in units of the central electron temperature, as shown in Fig. \ref{fig_boundary_potential} (a), then we find not only a fairly significant (30\%) difference from the theoretical prediction \eqref{eq_potential}, but also a small (5\%) difference between PIC simulations with different spatial grids. The reason for the first difference is the non-uniformity of the electron temperature observed in the simulations (Fig. \ref{fig_boundary_potential} (d)), which contradicts the assumption of the Debye sheath theory, where an isothermal ($T_e=$const) Boltzmann distribution is assumed for electrons.
\begin{figure*}
	\centering{\includegraphics[width=0.75\linewidth]{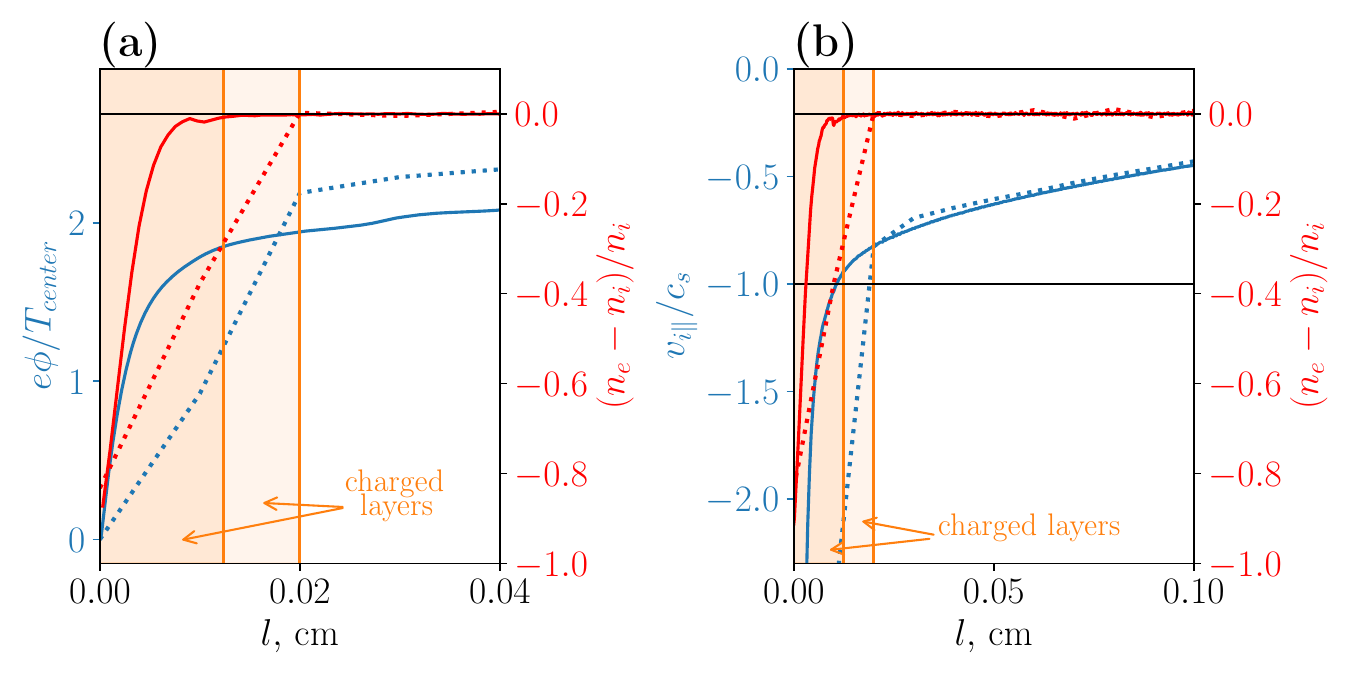}}
	\caption{Near-wall plasma parameters for PIC simulations with $h = 0.000145$ cm (solid curve) and $h = 0.01$ cm (dotted curve) in the steady state ($t=100$ ns): (a) potential in units of the electron temperature at the center of the system $T_{center}$; (b) longitudinal ion velocity in units of the speed of sound $c_s = \sqrt{T_{center} / m_i}$. Red curves on both plots correspond to the relative density difference $(n_e-n_i)/n_i$ and show where the quasi-neutrality is violated (orange stripes). All graphs are time averaged.}
	\label{fig_boundary_potential_w}
\end{figure*}
The second difference is explained by the fact that in simulations with different grids, different central electron temperatures are established, which is seen from the spatial profiles presented in Fig. \ref{fig_boundary_potential} (d). The difference in temperature indicates a difference in energy losses and is explained by the influence of numerical noise, which hinders the transport of slow particles while having little effect on the transport of fast particles. For this reason, the central electron temperature in the coarse-grid, higher-noise simulation is slightly lower than the temperature in the less noisy simulation with Debye-radius resolution. Despite the contradictory nature of the formulation of the problem in the theory and quantitative disagreements of 30\%, the model correctly reproduces the main features of the spatial structure of the sheath. In particular, in front of the charged sheath (Fig. \ref{fig_boundary_potential_w} (a)), a quasi-neutral pre-sheath is observed, in which the potential drop from the plasma center to the beginning of the near-wall charged layer is $e\Delta\phi/T_{center} = 0.7$ in the simulation with the resolution of the Debye radius and $e\Delta\phi/T_{center} = 0.6$ in the simulation with a coarse grid. This is somewhat higher than the well-known Bohm criterion, in which a jump of $e\Delta\phi/T_{center} = 0.5$ should ensure the acceleration of ions to the speed of sound at the entrance to the charged layer (see, for example, \cite{Lochte1968plasma}). However, as can be seen from the graphs of the average flux velocity of ions (Fig. \ref{fig_boundary_potential_w} (b)), the transition through the sonic threshold $v_{i\|}=c_s$ in both PIC simulations actually occurs at the boundary of the charged layer. Thus, the potential jump in the pre-sheath, as before, performs the function of accelerating ions to the average sonic velocity, and its higher value is explained by the fact that  a significant portion of the ions injected between the center of the system and the wall is affected by only a part of the full drop in accelerating potential.

Comparing the calculations with a fine and coarse grid, we can conclude that the model correctly reproduces the near-wall jump in electric potential even without resolving the fine structure of the Debye sheath.

\section{Modeling of longitudinal classical losses in mirror traps} \label{sec_open_trap}

To find out the role played by electron kinetics in the problem of plasma confinement in a mirror magnetic field, we consider formation of stationary plasma profiles in the presence of a constant source of particles in the central part of the trap and compare the results of fully kinetic simulations using the ADEPT code  with the results of hybrid simulations performed by the MIDAS code  \cite{prikhodko2025numerical}. In the hybrid code, a kinetic equation with a collision term is solved for ions, while a simplified fluid model is used for electrons. In this model, the electron distribution is assumed to be Maxwellian, their density is everywhere equal to the ion density, their temperature is assumed to be constant in space and determined by the balance of energy gained from the ions and lost to the walls. The longitudinal force balance in this massless fluid plays the role of equation for the electric potential. Moreover, solutions in the hybrid model are sought only between magnetic mirrors, while the expanders and walls are taken into account through specific boundary conditions \cite{prikhodko2025numerical}.
\begin{figure*}
	\centering{\includegraphics[width=\linewidth]{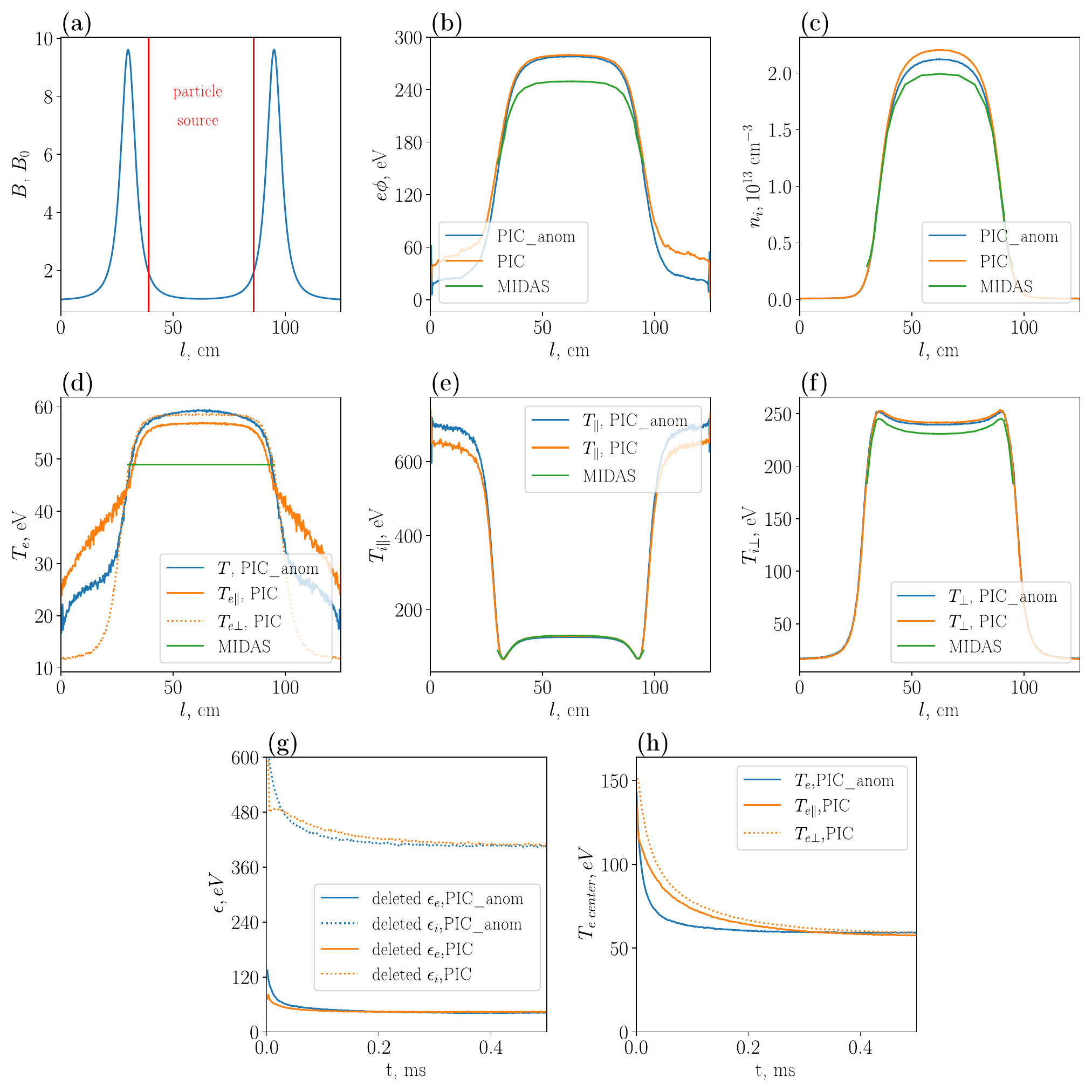}}
	\caption{Comparison of the results obtained in the PIC simulation with anomalous e-e collisionality (blue curves), PIC simulation with the real e-e collision rate (orange curves) and the hybrid MIDAS simulation (green curves): (a) magnetic field profile; (b) plasma potentials relative to the conducting wall; (c) ion density profiles; (d) - (f) average longitudinal $T_\parallel$ and transverse $T_\perp$ energies of electrons and ions; (g) average energy carried away by one particle; (h) electron temperature at the center of the system. All spatial profiles are given for a steady state, all graphs (except for the MIDAS curves) are time averaged. By $T_\parallel$ we mean the average doubled longitudinal kinetic energy $<mv^2_\parallel>$ of particles of a given type, by $T_\perp$ we mean the average transverse energy $<mv^2_\perp/2>$ of these same particles.}
	\label{fig_open_trap}
\end{figure*}
Since in the kinetic plasma confinement regime the Maxwellian nature of the distribution is not guaranteed even for electrons, the comparison with the hybrid model will be carried out not only in the case of real collisions between electrons, but also in a regime where the frequency of electron-electron collisions will be artificially increased. It will allow us to ensure the equilibrium of electrons and to fulfill at least one of the assumptions of the hybrid model.

Both models solve the problem in the following formulation: in a starting plasma uniformly occupying the space between mirrors with the initial density $n = 10^{13}$ cm$^{-3}$ and temperatures $T_{e0}=T_{i0}=150$, we inject electron-ion pairs with Maxwellian velocity distributions, the same temperatures and a constant rate of density increase $dn_{\rm inject}/dt = 1.17\cdot 10^{17}$ cm$^{-3}$/s. The injection region is located in the center of the trap, and its boundaries are shown in the figure \ref{fig_open_trap} (a). All types of collisions (e-e, e-i, i-i) are taken into account, and PIC simulations are carried out using both real electron collisionality and an anomalously high frequency of e-e collisions, which is implemented in the model by increasing the Coulomb logarithm by 4 orders of magnitude ($\Lambda_{\rm anomal}=10^4 \Lambda_{ee}$). Unlike previous simulations, in this problem the Coulomb logarithm is calculated for each pair of particles based on their individual relative velocity $u$
\begin{equation}
	\Lambda_{s_1 s_2}=\ln\left|\frac{\lambda_D  m u^2}{q_{s_1} q_{s_2}}\right|,
\end{equation}
where $\lambda_D$ is the Debye radius calculated from the average particle density and temperature in the cell, $m=m_{s_1}m_{s_2}/(m_{s_1}+m_{s_2})$. In PIC simulations, a unit density (chosen as $n=10^{13}$ cm$^-3$) corresponds to 500 particles in the cell at the magnetic mirror location and $\approx$ 5000 at the center of the trap.

The results of all simulations are presented in Fig. \ref{fig_open_trap}. Let us first compare the PIC simulations with the real (orange curves) and anomalous (blue curves) e-e collision frequencies and assess the significance of electron nonequilibrium in the problem. It can be seen that there is a small difference in the plasma and electric potential profiles between the magnetic mirrors. This is due to the large jump in the ambipolar potential between the plasma center and the wall ($e\Delta\phi\sim 300$ eV), which corresponds to 4-5 electron temperatures at the plasma center $T_{e\;center}$ (graphs (b), (d)). This potential drop confines most of the electrons within the open trap, thereby allowing them to undergo Maxwellization. However, in the region of the expander, where there is no magnetic field confinement and the density is low, all plasma components can be considered collisionless. This leads to anisotropy of the electron distribution function and a strong local difference in the longitudinal and transverse electron temperature profiles (Fig. \ref{fig_open_trap} (d)). This anisotropy arises due to the conservation of the electron magnetic moment, which causes almost all of their transverse energy to be transformed into longitudinal energy as they flow out of the magnetic mirror. An increase in the longitudinal temperature of electrons in expanders in the case of real collisions leads to a slower drop in the potential between the mirror and the wall (Fig. \ref{fig_open_trap} (b)) and a less intense acceleration of ions in this gap, which can be seen from the profile of the total longitudinal energy of ions (Fig. \ref{fig_open_trap} (e)) including the energy of directed motion.

Comparison with the results obtained by the MIDAS code (green curves) shows that the hybrid simulations reproduce the same longitudinal and transverse ion energy profiles, but predict a 15\%-20\% lower electron temperature, lower plasma potential and lower density. We attribute these differences to a violation of the hybrid model's assumption of isothermal Boltzmann electron distribution, since in the PIC simulation the electron temperature drops sharply in the expanders.

Based on PIC simulations, we can also determine the energy carried out of the trap by an electron-ion pair. Dividing the energy of  particles removed over several time steps by their number, we find that, in a steady state, each escaping ion has an energy of $\approx 6.5T_{e\;center}$, and each electron has an energy of $T_{e\;center}$ (Fig. \ref{fig_open_trap} (g), (h)). Thus, the energy loss per electron-ion pair is $\approx 7.5 T_{e\;center}$, which is consistent with experimental measurements at the GDT facility \cite{soldatkina2020measurements}.

The plasma lifetime in PIC simulations is determined by the steady-state value of the density $n_{\infty}$:
\begin{equation}
	\tau_{\rm simul} = \frac{n_{\infty}}{dn_{\rm inject}/dt} = 185\; \text{$\mu$s}.
\end{equation}
Since the effective path length of the ion turns out to be significantly greater than the distance between the plugs $L$
\begin{equation}
	\frac{\lambda_{\rm eff}}{L}=\frac{v_{Ti}\tau_{ii}}{L}\frac{\ln R}{R}\sim 10,
\end{equation}
our simulations reproduce the kinetic confinement regime (here $R=B_{plug}/B_{center}=9.4$ is the mirror ratio, $\tau_{ii}$ is the time of ion-ion collisions calculated based on the average temperature
\[T_i = \frac{2}{3}\left(\frac{T_{i\parallel}}{2} + T_{i\perp}\right) = 202\;\text{eV},\]
 $v_{Ti}=\sqrt{T_{i\|}/m_i}$ is the average longitudinal ion velocity, the average Coulomb logarithm is estimated as $\Lambda=13$). A rough theoretical estimate for the plasma lifetime in this regime (not accounting for the ambipolar effect) is given by the Budker formula
\begin{equation}
	\tau_{\rm theory}^B = 0.4 \tau_{ii} \ln R = 0.4\frac{\sqrt{2}}{\pi} \frac{\sqrt{m_i T_i^3}}{e^4 \Lambda n_i} \ln R \approx 200\,\mbox{\rm $\mu$s}
\end{equation}
and agrees well with the simulation result. In a steady state, the same retention time should be achieved for electrons. The well-known formula of Pastukhov, generalized by Cohen \cite{Cohen1978}, 
\begin{widetext}
\begin{equation}
	\tau_{\rm theory}^{PC}=\frac{\sqrt{\pi}}{4}\tau_{ee} \sqrt{1+\frac{1}{R}}\ln\left(\frac{\sqrt{1+1/R}+1}{\sqrt{1+1/R}-1}\right) \frac{\xi e^\xi}{\displaystyle 1+\frac{e^\xi}{2}\sqrt{\frac{\pi}{\xi}}{\rm erfc}\sqrt{\xi}}
\end{equation}
for the observed potential jump $\xi=e\Delta\phi/T_e\approx 4.8$, overestimates the retention time by several times $\tau_{\rm theory}^{PC}\approx 746$ $\mu$s. The alternative Chernin-Rosenbluth-Cohen formula \cite{Cohen1978}
\begin{equation}
	\tau_{\rm theory}^{CRC}=\frac{\sqrt{\pi}}{4}\tau_{ee} \xi e^\xi \frac{\left[\ln\left(4R\xi/\delta\right)-2\right]^2}{\ln\left(4R\xi/\delta\right)-2+\ln\left(0.85\xi\right)}, \qquad \delta=\sqrt{\frac{\delta_p}{\ln(R\xi)}},
\end{equation}
where $\delta_p$ is a solution of the equation  
\begin{equation}
	\delta_p=\ln\left\{1-\delta_p\left(1+\frac{\delta_p}{\xi}\right)\left[1-\xi\ln\left(\frac{e^2\delta_p}{4R(\xi+\delta_p)}\right)\right]\ln\left(\frac{e^2\delta_p}{4R(\xi+\delta_p)}\right)\right\},
\end{equation}
\end{widetext}
reduces the retention time to the value $\tau_{\rm theory}^{CRC}\approx 428$ $\mu$s, but also does not provide good quantitative agreement with the modeling. Thus, theoretical estimates claiming to describe the ambipolar effect predict particle lifetimes only to an order of magnitude.

\section{Summary}

This paper presents an extended version of the 1D2V semi-implicit drift-kinetic PIC code ADEPT, which incorporates the Takizuka-Abe binary Coulomb collision algorithm and implements new boundary conditions in the form of perfectly conducting walls with a floating potential. In this form, the model can be used to study classical longitudinal losses from open traps.
To verify the correctness of the collision module, we simulated the problem of relaxation to equilibrium in a plasma with different ion and electron temperatures and showed good quantitative agreement with the analytical theory  if the number of macroparticles per cell exceeds 5000. It is shown that the boundary conditions requiring the absorption of particles at the walls and the zeroing of the electric field inside them can be realized without violating the law of energy conservation, and the jump of the ambipolar potential in the near-wall Debye layer is correctly reproduced even without the resolution of the Debye scale. This makes it possible to  use a large grid in drift-kinetic PIC simulations without fear of markedly affecting the electron confinement.

To demonstrate the importance of taking electron kinetics into account in the problem of plasma confinement in an open trap, steady-state profiles of various plasma parameters were calculated in the presence of a constant Maxwellian particle source. A comparison of the fully kinetic PIC simulations with the hybrid ones (kinetic ions --- fluid electrons)  revealed significant differences (at the 15\% level) in the electron temperature, potential jump, and confined plasma density. The reason for this is that the concept of the isothermal Boltzmann electrons does not work in expanders.

\begin{acknowledgments}
The work is supported by the Russian Science Foundation (project 24-12-00309).

Simulations were performed using the computing resources of the Information and Computing Center in Novosibirsk State University.
\end{acknowledgments}

\bibliography{ref}

\end{document}